\documentclass[final,1p,times]{elsarticle}
\usepackage{graphics}
\usepackage{graphicx}
\usepackage{epstopdf}
\usepackage{ifpdf}
\usepackage{amssymb}
\usepackage{amsthm}
\usepackage{lineno}
\usepackage{dcolumn}
\usepackage{bm}
\usepackage{epsfig}
\usepackage{ifpdf}
\usepackage{amsthm}
\usepackage{subfig}

\newdefinition{rmk}{Remark}
\newproof{pf}{Proof}
\newproof{pot}{Proof of Theorem \ref{thm2}}

\begin{document}
\begin{frontmatter}
\title{Scalar field scattering by a Lifshitz black hole under a non-minimal coupling}
\author[ucv]{Samuel Lepe}
\ead{slepe@ucv.cl}
\author[dcf]{Javier Lorca}
\ead{j.lorca@ufro.cl}
\author[dcf]{Francisco Pe\~na}
\ead{fcampos@ufro.cl}
\author[dcf]{Yerko V\'asquez}
\ead{yvasquez@ufro.cl}
\address[ucv]{Instituto de F\'\i sica, Facultad de Ciencias, Pontificia Universidad Cat\'olica de Valpara\'\i so, Casilla 4059, Valpara\'\i so, Chile}
\address[dcf]{Departamento de Ciencias F\'\i sicas, Facultad de Ingenier\'\i a, Ciencias y
Administraci\'on, Universidad de La Frontera, Avda. Francisco Salazar 01145,
Casilla 54-D Temuco, Chile.}

\begin{abstract}
We study the behavior of a scalar field under a $z = 3$ Lifshitz black hole background, in a way that is non-minimally coupled to the gravitational field. A general analytical solution is obtained along with two sets of quasinormal modes associated to different boundary conditions that can be imposed on the scalar field, non-minimal coupling parameter appears explicitly on these solutions. Stability of quasinormal modes can be studied and ensured for both cases. Also, the reflection and absorption coefficients are calculated, as well as the absorption cross section which features an interesting behavior because of being attenuated by terms strongly dependant on the non-minimal coupling. By a suitable change of variables a soliton solution can also be obtained and the stability of the quasinormal modes are studied and ensured.
\end{abstract}

\end{frontmatter}

\linenumbers

\section{Introduction}

It is known that theories such as New Massive Gravity (NMG) admits Lifshitz black holes as solutions, whose particularity is that they are invariant under an anisotropic scale transformation of the form $t \rightarrow \lambda^{z}t$  y  $x \rightarrow \lambda x$, where $z$  is called the relative scale between time and space dimensions, specifically $z=3$ for the aforementioned theory. These black hole solutions have come to prominence because they might provide a way to extend the AdS/CFT correspondence \cite{Maldacena} to systems found on non-relativistic condensed matter physics which features a very similar behavior \cite{Son,Hartnoll1,Hartnoll2}, where it was proven that the relaxation time of thermal states of the conformal theory at the boundary is proportional to the inverse of the imaginary part of the quasinormal modes of the dual gravity background \cite{Horowitz}.

As a way to consider explicitly an interaction between gravity and a scalar field, a non-minimal coupling is added on the equation of motion for the scalar field, motivated in resemblance of those found in theories in which the action has this type of coupling \cite{Buchbinder,Elizalde,Muta,Lepe}. As a consequence, the Ricci scalar appears directly in the equation of motion; however, in a Lifshitz background the Ricci scalar has a relatively simple form which allows to obtain an analytical solution.

In this paper we focus on the study of the reflection and absorption coefficients, as well as the absorption cross section \cite{Birmingham,Kim,Harmark,Gonzalez}, some efforts have been made for the minimally coupled Lifshitz black hole \cite{Moon} in this direction; also, it has been shown that for spherically symmetric black hole and a massless minimally coupled scalar field the cross section equals the area of the horizon \cite{Das}; however, we will show that there is a strong dependence on the non-minimal coupling parameter in this case, this imply that the absorption cross section is also dependent on it, hence not allowing to obtain as result the geometric area of the black hole in the limit of low energies, unless the non-minimal coupling became null.

Gravitational waves predicts a non normal type of oscillation mode where the frequencies become complex or also called quasinormal, with the real part representing the frequency of oscillation and the imaginary part representing the damping \cite{Kokkotas}. This study has already been done to some of this black hole solutions \cite{Bertha,yerkillo,Julio,Myung} considering a scalar field moving over a Lifshitz background, where no imaginary parts have been found so far. The presence of the non-minimal coupling does not affect this behavior thanks to the simple form of the Ricci scalar. By relaxing the boundary conditions used, allowing to be Dirichlet and Neumann mixed, one can obtain a new set of quasinormal modes previously not found which can be analyzed to study their stability. On the other hand, by performing two Wick's rotations between time and space coordinates on the metric it is possible to find a soliton solution \cite{soliton}, find its quasinormal modes and study its stability.

This paper is organized as follows, the first section introduces the formalism, the field equations to be used and generalities on the Lifshitz metric for $z = 3$, in the second section we find the solution of the Klein-Gordon differential equation for a scalar field on this Lifshitz space-time and formally treats the suitable boundary conditions to this solution, third section find the reflection and absorption coefficients along with the cross section, which features and interesting dependency on the non-minimal coupling parameter, fifth and sixth section are committed to the study of the quasinormal modes their stability for the black hole and its related soliton solution. Finally, we stress out the important results of this paper as final remarks.

\section{Formalism and field equations}
Let us consider the typical NMG action
\begin{eqnarray}  \label{action}
\mathcal{S}(g)=\int d^{3}x \sqrt{-g}\mathcal{L},
\end{eqnarray}
where
\begin{equation}  \label{lagrangiana}
\mathcal{L}=\frac{{1}}{{2}}R -2 \Lambda_0 - \frac{1}{\nu^2} \left( R^{\mu \nu} R_{\mu \nu} -\frac{3}{8} \, R^2 \right),
\end{equation}
in this Lagrangian density we use natural units, i. e. $8 \pi G = c = 1$. $R$ denotes the scalar of curvature, $R^{\mu \nu}$ denotes the Ricci tensor. In order to ensure a complete correspondence to gravitational solutions the parameters $\nu$ and $\Lambda_0$ must be chosen to be $\nu^2 = - \frac{1}{2 l^2}$ (with mass dimension) and $\Lambda_0 = -\frac{13}{2 \, l^2}$ (the cosmological constant) respectively.

The field equations are obtained by varying the total action (\ref{action}) with respect to the metric. We will use the 3-dimensional Lifshitz black hole background as a known solution from this theory \cite{Ayon}
\begin{equation}\label{lipshitzmetric}
ds^2= - \left( \frac{\rho}{l} \right)^4 f^2\left(\rho\right) dt^2 + \frac{1}{f^2\left(\rho\right)} d\rho^2 + \rho^2 d\varphi^2,
\end{equation}
with $f^2 \left( \rho \right) = \left( \frac{\rho}{l} \right)^2 - M$, where $l$ is the curvature radius of the Lifshitz space-time and $M$ represents the black hole mass. From this final form the Ricci scalar (for $ z = 3 $) can be calculated to be
\begin{equation}\label{Ricciscalar}
R = 8 \, \frac{M}{\rho^2} - \frac{26}{l^2}.
\end{equation}
Let us consider a massive scalar field on this background, in a way it is weakly coupled to the gravitational field, in the sense that the presence of this field does not perturb the background metric, but it is just a field that moves along this geometry. A typical equation followed by this scalar field is
\begin{eqnarray}
\label{phiequations} \left( \square - m^2 \right) \phi & =& \frac{d \gamma \left( \phi \right)}{d \phi} R - \frac{d U \left(\phi \right)}{d\phi },
\end{eqnarray}
where $\gamma =\frac{1}{2} \left( 1-\xi \phi ^{2} \right)$ and $U\left(\phi\right)$ is the self-interacting potential density. As was already mentioned, this type of equation is motivated by the kind of field equation obtained when varying an action with a scalar field with non-minimal coupling parameter $\xi$ to the Ricci Scalar \cite{Lepe}. For finiteness, consider a null self-interacting potential density (i.e. $U \left( \phi \right) = 0$, equation (\ref{phiequations}) takes the form
\begin{eqnarray}\label{nonlinearKleinGordon}
\left( \square - m^2 \right) \phi = -\xi \phi R \, .
\end{eqnarray}

\section{Solution of the differential equation}

Starting from equation (\ref{nonlinearKleinGordon}) and using the metric (\ref{lipshitzmetric}) it can be shown that a separation of variables for the field $\phi$ of the form
\begin{equation}\label{separationofvariables}
\phi \left( \rho , \varphi , t \right) = \mathcal{R}\left( \rho \right) \exp \left(-i \left( \, \omega \, t + \, \kappa \, \varphi \right) \right),
\end{equation}
allows to write the radial equation in the following form
\begin{equation}\label{radialdiffrho}
\partial_{\rho}^2 \mathcal{R} \left( \rho \right) + \frac{\left( 5 \frac{\rho}{l^2} - 3 \frac{M}{\rho} \right)}{\left( \frac{\rho^2}{l^2} - M \right)} \partial_{\rho} \mathcal{R} \left( \rho \right) + \frac{1}{\left(\frac{\rho^2}{l^2} - M \right)} \left[ \frac{l^4 \omega^2}{\rho^4 \left(\frac{\rho^2}{l^2} - M \right)} - \frac{\kappa^2}{\rho^2} - m^2 + \xi \left(\frac{8M}{\rho^2} - \frac{26}{l^2}\right) \right] \mathcal{R} \left( \rho \right) = 0.
\end{equation}
By noting that the horizon of the black hole is located at $\rho_{+} = l \sqrt{M}$, let us define the scaling variable
\begin{equation}\label{scalingx}
x = 1 - \left(\frac{ \rho_+}{\rho}\right)^2 \,
\end{equation}
so that equation (\ref{radialdiffrho}) can be written in the form
\begin{equation}\label{radialdiffx}
\mathcal{R} '' \left( x \right) + \frac{1}{x \left( 1 - x \right)} \mathcal{R} ' \left( x \right) + \frac{1}{4 x \left( 1 - x \right)^2}  \left[ \frac{l^2 \omega^2 \left( 1 - x \right)^3}{M^3 x} - \frac{\kappa^2 \left( 1 - x \right)}{M} - l^2 m^2 + \xi \left( 8 \left( 1 - x \right) - 26 \right) \right] \mathcal{R} \left( x \right) = 0 \, .
\end{equation}
Equation (\ref{radialdiffx}) is of the Fuchsian type with two regular singular points located at $x=0$ and $x=1$, and one irregular singular point located at $x = - \infty $, which tell us that this equation is somewhat related to the confluent Heun family of equations. In fact, the following transformation $ \mathcal{R} \left( x \right) = x^{\alpha} \left( 1 - x \right)^{\beta} \mathcal{F} \left( x \right)$, justified by Fuch's theorem around the regular singular points, allows to identify this assertion better, yielding
\begin{equation}\label{frobenius}
x \left( x - 1 \right) \mathcal{F} '' \left( x \right) + \left[ \left( b + 1 \right) \left( x - 1 \right) + \left( c + 1 \right) x \right] \mathcal{F} ' \left( x \right) + \left( d x - \epsilon \right) \mathcal{F} \left( x \right) = 0 \, ,
\end{equation}
where
\begin{eqnarray}
\label{alphaheun} \alpha_\pm & = & \pm \frac{i}{2 M^{3/2}} \omega l \, , \\
\label{betaheun} \beta_\pm & = & 1 \pm \sqrt{1 + 13 \frac{\xi}{2} + \frac{m^2 l^2}{4} } \, ,\\
\label{bheun} b & = & 2 \alpha_\pm \, , \\
\label{cheun} c & = & 2 \beta_\pm - 2 \, , \\
\label{dheun} d & = & - \frac{l^2 \omega^2}{4 M^3} \, ,
\end{eqnarray}
and
\begin{equation}\label{eheun}
\epsilon = \alpha_\pm + \beta_\pm - \left( \alpha_\pm + \beta_\pm \right)^2 - \frac{\kappa^2}{4 M} - \frac{l^2 \omega^2}{2 M^3} + 4 \frac{\xi}{2} .
\end{equation}
Using the previous parameters, the solution of equation (\ref{frobenius}) is written
\begin{eqnarray}
\label{solutionfrobenius} \mathcal{F} \left( x \right) & = & C_1 \, \mathrm{Heun_C} \left( 0 , b , c , d , - \frac{1}{2} \{ 1 + c \}b - \frac{c}{2} - \epsilon , x \right) \\
\nonumber & & + C_2 \, x^{-b} \, \mathrm{Heun_C}\left( 0 , - b , c , d , - \frac{1}{2} \{ 1 + c \}b - \frac{c}{2} - \epsilon , x \right),
\end{eqnarray}
where $\mathrm{Heun_C}$ are the confluent Heun functions. Finally, the solution of equation (\ref{radialdiffx}) is
\begin{eqnarray}
\label{solutionradialdiffx} \mathcal{R} \left( x \right) & = & C_1 \, x^{\alpha} \, \left( 1 - x \right)^{\beta}  \mathrm{Heun_C} \left( 0 , b , c , d , - \frac{1}{2} \{ 1 + c \}b - \frac{c}{2} - \epsilon , x \right) \\
\nonumber & & + C_2 \, x^{- \alpha} \, \left( 1 - x \right)^{\beta} \, \mathrm{Heun_C}\left( 0 , - b , c , d , - \frac{1}{2} \{ 1 + c \}b - \frac{c}{2} - \epsilon , x \right) \, .
\end{eqnarray}

Note that the simplicity of the Ricci scalar for this background allows to identify directly from equation (\ref{radialdiffrho}) the following transformations from the homogeneous problem on \cite{Myung}
\begin{equation} \label{transformationfrombertha}
\kappa^2 \rightarrow \kappa^2 - 8 M \xi \;\;\;\;\; m^2 \rightarrow m^2 + 26\frac{\xi}{l^2} \, ,
\end{equation}
which means that the non-minimal coupling problem is formally equivalent to the homogeneous one. This transformation is useful when calculating the quasinormal modes of this black hole.

\subsection{Asymptotic Expressions.}

In order to obtain the reflection and absorption coefficients, the asymptotic expressions for this solution must be obtained. To incorporate boundary conditions will force us to focus on two distinct points

\begin{itemize}
\item $\rho \rightarrow \rho_+ \Rightarrow x \rightarrow 0$.

Here the solution takes the following approximated form
\begin{eqnarray}
\nonumber \mathcal{R} \left( x \rightarrow 0 \right) & \approx & C_1 x^{\alpha} + C_2 x^{- \alpha} \\
\label{boundaryhorizon} & = & C_1 \exp \left( {\alpha} \ln x \right) + C_2 \exp \left( {- \alpha} \ln x \right) \, .
\end{eqnarray}

Let us recall that there exists two values for $\alpha$, and so, let us also assume that $\alpha = \alpha_- = \frac{- i \omega l}{2 M^{3/2}}$, then by equation (\ref{boundaryhorizon}) the condition $C_2 = 0$ arises by considering just ingoing flux in the horizon. We recall that choosing $\alpha = \alpha_+$ will derive in the same result because of considering the same flux conditions, which will lead to set the constant $C_1=0$.

\item $\rho \rightarrow \infty \Rightarrow x \rightarrow 1$.

This case is a bit more subtle, here we have to use the following identities regarding the confluent Heun functions \cite{Kwon}:
\begin{eqnarray}
\label{HeunCidentity1} \mathrm{Heun_C} \left( 0, b, c, d, e, 0 \right) &=& 1 \\
\label{HeunCidentity2} \mathrm{Heun_C} \left( 0 , b , c , d , e ; x \right) & = & D_1 \, \frac{ \Gamma \left( b + 1 \right) \Gamma \left( - c \right)}{ \Gamma \left( 1 - c + k \right) \Gamma \left( b - k \right)} \, \mathrm{Heun_C} \left( 0 , c , b , - d , e + d ; 1 - x \right) \\
\nonumber &+& D_2 \, \left( 1 - x \right)^{-c} \frac{ \Gamma \left( b +1 \right) \Gamma \left( c \right)}{ \Gamma \left( 1 + c + s \right) \Gamma \left( b - s \right)} \, \mathrm{Heun_C} \left( 0 , - c , b , - d , e + d ; 1 - x \right) \, ,
\end{eqnarray}
where the following set of equations are found to be satisfied between the parameters of the confluent Heun's functions
\begin{eqnarray}
\label{k} k^2 + \left( 1 - b - c \right) k - \epsilon - b - c +  \frac{d}{2} & = & 0 \, , \\
\label{s} s^2 + \left( 1 - b + c \right) s - \epsilon - b \, \left( c + 1 \right) + \frac{d}{2} & = & 0 \, , \\
\label{e}  - \frac{1}{2} \left( 1 + c \right) b - \frac{c}{2} - \epsilon & = & e \, .
\end{eqnarray}
Therefore, equation (\ref{solutionradialdiffx}) takes the following asymptotic form on infinity
\begin{eqnarray}
\nonumber \mathcal{R} \left( x \rightarrow 1 \right) \approx C_1 \left( 1 - x\right)^{\beta} \left[B_1 + B_2 \left( 1 - x \right)^{2 - 2 \beta} \right] + C_2 \left( 1 - x \right)^{\beta} \left[B_3 + B_4 \left( 1 - x \right)^{2 - 2 \beta} \right] \, ,
\end{eqnarray}
however, we have already discarded the constant $C_2$ by flux conditions on the horizon, hence the appropriate asymptotic expression to be used is
\begin{equation}\label{boundaryinfinity}
\mathcal{R} \left( x \rightarrow 1 \right) \approx C_1 \left( 1 - x\right)^{\beta} \left[B_1 + B_2 \left( 1 - x \right)^{2 - 2 \beta} \right] \, ,
\end{equation}
where
\begin{eqnarray}
\label{b1} B_1 & = & D_1 \frac{ \Gamma \left( b + 1 \right) \Gamma \left( - c \right)}{\Gamma \left( 1 - c + k \right) \Gamma \left( b - k \right)} \, ,\\
\label{b2} B_2 & = & D_2 \frac{ \Gamma \left( b + 1 \right) \Gamma \left( c \right)}{\Gamma \left( 1 + c + s \right) \Gamma \left( b - s \right)} \, .
\end{eqnarray}
\end{itemize}

As a way to check the solution of equation (\ref{boundaryinfinity}) the asymptotic equation on the infinity will be solved. This can be done by performing in (\ref{radialdiffx}) the following change of variable $y = 1 - x$  with $y \rightarrow 0$, hence it follows
\begin{equation}\label{radialdiffasymptotic}
y^2 \frac{d^2 R \left( y \right)}{dy^2} - y \frac{d R \left( y \right)}{dy} + \left[ E_1 + E_2 y \right]R \left( y \right) = 0 \, ,
\end{equation}
where we have only retain terms up to a second order in $y$ and $E_1 =  \frac{-1}{4} \left[\left( lm \right)^2 + 26 \xi \right]$ and $E_2 = \frac{1}4{} \left[8 \xi - \frac{\kappa^2}{M} \right]$. The solution to this equation can be written
\begin{eqnarray*}
R \left( y \rightarrow 0 \right) &=& y \{ F_1 \, E_2 \, \Gamma \left(1 - 2 \sqrt{1 - E_1}\right)  \mathrm{J}_{-\nu} \left( u \right)\\
 & & + F_2 \, E_2 \, \Gamma \left( 1 + 2 \sqrt{1 - E_1}  \right) \mathrm{J}_{\nu} \left( u \right) \}
\end{eqnarray*}
where $\mathrm{J}_\nu \left( u \right)$ are the Bessel's functions of the first kind, having $\nu = {2 \sqrt{ 1 - E_1}}$ and $u = 2 \sqrt{E_2 y} $. Let us use the following expansion of the Bessel's functions for $u \ll 1$
\begin{equation} \label{Besselaaproximation}
\mathrm{J}_\nu \left( u \right) = \frac{u^{\nu}}{2^{\nu} \Gamma \left( 1 + \nu \right)} \{ 1 - \frac{u^2}{2 \, \left( 2 \nu + 2 \right)} + \mathcal{O} \left( u^4 \right) \},
\end{equation}
and by restricting the previous expression up to the first term, the solution for the asymptotic radial equation can be written in the form
\begin{equation}\label{radialasymptoticsolution1}
R \left( y \rightarrow 0 \right) = F_1 \, E_2^{1 -\sqrt{1 - E_1}} \, y^{1 - \sqrt{1 - E_1}} + F_2 \, E_2^{1 + \sqrt{1 - E_1}} \, y^{1 + \sqrt{1 - E_1}}.
\end{equation}

Note that $1 -\sqrt{1 - E_1} = \beta_{-}$, so equation (\ref{radialasymptoticsolution1}) and (\ref{boundaryinfinity}) are equivalent, therefore the corresponding coefficients must be equal
\begin{eqnarray}
\label{hatF1} \hat F_1 =  F_1 \, E_2^{1 -\sqrt{1 - E_1}} = C_1 \, B_1 \, , \\
\label{hatF2} \hat F_2 =  F_2 \, E_2^{1 +\sqrt{1 - E_1}} = C_1 \, B_2 \, .
\end{eqnarray}

\section{Reflection and Absorption coefficients}

Before going any further, it is convenient to express the equations needed explicitly on the scaling variable $x$ (equation (\ref{scalingx})). The flux $F$ is known to be defined by \cite{Satoh,Gonzalez}:
\begin{eqnarray}
\nonumber F & = & \frac{\sqrt{-g}g^{\rho \rho}}{2i} \left( R^* \left( \rho \right) \partial_\rho R \left( \rho \right) - R \left( \rho \right) \partial_\rho R^* \left( \rho \right) \right) \, ,   \\
\label{F} & = & 2 \, \frac{{\rho_+}^4 }{l^4} \left[ \frac{ x }{ 1 - x } \right] \textit{Im} \{ R^* \left( x \right) \partial_x R \left( x \right)\},
\end{eqnarray}
and by using equations (\ref{radialasymptoticsolution1}), (\ref{hatF1}) and (\ref{hatF2}) the following expression is obtained
\begin{equation}\label{asymptoticflux1}
F_{asymptotic} = \frac{4\,\left(\beta - 1\right)}{l^4} \textit{Im} \{\hat F_1 \, \left({\rho_{+}}^4 \hat F_2^* \right) \}.
\end{equation}
As it has been discussed by other authors the problem that equation (\ref{asymptoticflux1}) has is that it is impossible to determine wether the flux is ingoing or outgoing, however it can be written in the following form \cite{Birmingham,Kim,Oh}
\begin{equation}\label{asymptoticflux2}
F_{asymptotic} = \frac{4\,h\,\left(\beta - 1 \right)}{l^4} \left| \hat F_{asymptotic}^{in} \right|^2 - \frac{4\,h\,\left(\beta - 1 \right)}{l^4} \left| \hat F_{asymptotic}^{out} \right|^2,
\end{equation}
where
\begin{eqnarray}
\label{hatFin} \hat F_{asymptotic}^{in} = \frac{1}{2} \left( \hat F_1 + \frac{i \, {\rho_+^4} }{h} \hat F_2 \right) \, , \\
\label{hatFout} \hat F_{asymptotic}^{out} = \frac{1}{2} \left( \hat F_1 - \frac{i \, {\rho_+^4} }{h} \hat F_2 \right),
\end{eqnarray}
where $h$ is a real parameter with dimension $[L]^4$. Note that equation (\ref{asymptoticflux2}) coincides with the expression (\ref{asymptoticflux3}), this last one obtained by a different procedure.

On the other hand, using equation (\ref{boundaryhorizon}), on the horizon the following relation results
\begin{eqnarray}
\nonumber F_{horizon} & = & 2 \, \frac{{\rho_+}^4 }{l^4} \lim_{x \rightarrow 0} \left[ \frac{ x }{ 1 - x } \right] \textit{Im} \{ R^* \left( x \right) \partial_x R \left( x \right)\} \, ,\\
\nonumber & = & - \, \frac{{\rho_+}^4 }{l^4}|C_1|^2 \frac{\omega \, l}{M^{3/2}} \, ,\\
\label{horizonflux} & = & - \, \rho_+ \, \omega |C_1|^2 \, ,
\end{eqnarray}
where in the last line, the definition of the black hole mass has been used.

By using equations (\ref{hatFin}), (\ref{hatFout}), (\ref{hatF1}) and (\ref{hatF2}) we can calculate the reflection and absorption coefficients as
\begin{eqnarray}
\label{Reflection} \mathfrak{R} \equiv \left| \frac{F^{out}_{asymptotic}}{F^{in}_{asymptotic}} \right| & = & \frac{\left| B_1 \right|^2 + \frac{\rho_+^8}{h^2}\left| B_2 \right|^2 + \frac{2 \rho_+^4}{h}\mathfrak{Im}\left(B_1^* B_2 \right)}{\left| B_1 \right|^2 + \frac{\rho_+^8}{h^2}\left| B_2 \right|^2 - \frac{2 \, \rho_+^4}{h}\mathfrak{Im}\left(B_1^* B_2 \right)} , \\
\nonumber \mathfrak{U} \equiv \left| \frac{F^{in}_{horizon}}{F^{in}_{asymptotic}} \right| & = & \frac{ \omega \, l}{ M^{3/2} \, \left| h \right| \, \left| \beta - 1 \right| \, \left( \left| B_1 \right|^2 + \frac{\rho_+^8}{h^2}\left| B_2 \right|^2 - \frac{2 \, \rho_+^4}{h}\mathfrak{Im}\left(B_1^* B_2 \right) \right)} \, ,\\
\label{Absorption} & = & \frac{ \omega \, l^4 {\rho_+}}{ \left| h \right| \, \left| \beta - 1 \right| \, \left( \left| B_1 \right|^2 + \frac{\rho_+^8}{h^2}\left| B_2 \right|^2 - \frac{2 \, \rho_+^4}{h}\mathfrak{Im}\left(B_1^* B_2 \right) \right)} \, ,
\end{eqnarray}
where as in equation (\ref{horizonflux}), the definition of the black hole mass has been used again. As a manner to avoid any divergence on (\ref{Reflection}) and (\ref{Absorption}) we will choose negative values for $h$.

Recall that factors $B_1$ and $B_2$ are dependent on Gamma functions (equations (\ref{b1}) and (\ref{b2})) which makes difficult to work with the analytical expressions, however, the absorption cross section is immediate from equation (\ref{Absorption})
\begin{equation}\label{crosssection}
\sigma = \frac{\mathfrak{U}}{\omega} = \frac{l^4 {\rho_+}}{ \left| h \right| \, \left| \beta - 1 \right| \, \left( \left| B_1 \right|^2 + \frac{\rho_+^8}{h^2}\left| B_2 \right|^2 - \frac{2 \, \rho_+^4}{h}\mathfrak{Im}\left(B_1^* B_2 \right) \right)} \, .
\end{equation}

It is straightforward to verify that in the low energy limit for a s-wave type of solution ($\kappa = 0$), an expression for the absorption cross section can be obtained to be
\begin{equation}\label{koreanlimit}
\sigma \left( \xi \right) = \frac{1}{\sqrt{1+13\frac{\xi}{2}}}\frac{1}{\left| f \left( \kappa, \xi \right) \right|^2} 2 \pi \rho_+ < 2 \pi \rho_+ \, ,
\end{equation}
where we have chosen $\left| h \right| = \frac{l^4}{2 \pi}$ and $\left| f \left( \kappa, \xi \right) \right|^2$ is the term appearing on the denominator of equation (\ref{crosssection}) dependent on the coefficients $B_1$, $B_2$ and $h$. The presence of the $\xi$-parameter decreases the final value of the absorption cross section away from the geometric area even if we try the low energy limit, i. e. when $ m \rightarrow 0$, $\omega \rightarrow 0$. Equation (\ref{koreanlimit}) is reduced to the geometric area of the black hole $\sigma = 2 \pi  \rho_+$ only when $\xi = 0$ as expected \cite{Moon}.

\begin{figure}[h]
\centering
  \framebox{\includegraphics[width=4in,height=3in]{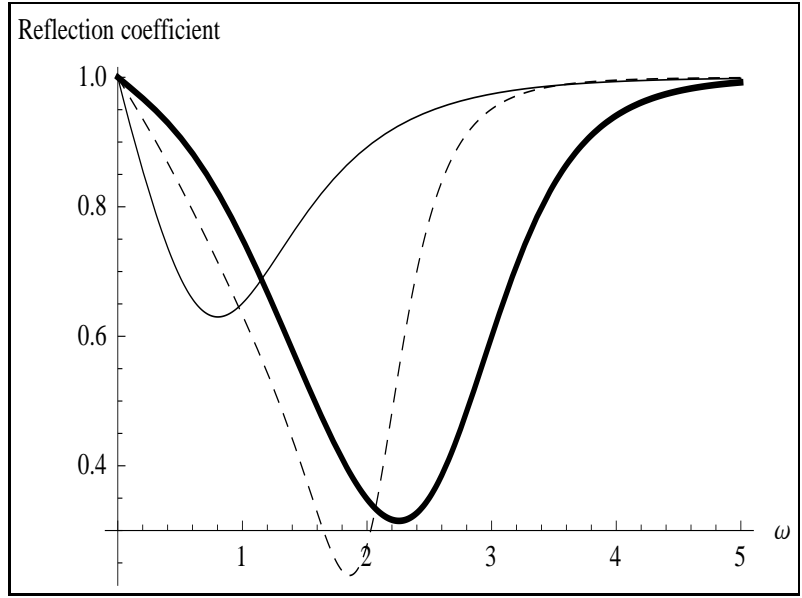}}\\
  \caption{\label{R} The behavior of the reflection coefficient with respect to $\omega$ for the following parameters $ M = 1; \; m^{2}l^{2}=-\frac{4}{5}; \; l=1; \; \kappa=1; h=-1$, for three different values $\xi=0$ (fine solid line); $\xi=0.1$ (dashed line) and $\xi=2$ (thick solid line)}
\end{figure}

\begin{figure}[h]
\centering
  \framebox{\includegraphics[width=4in,height=3in]{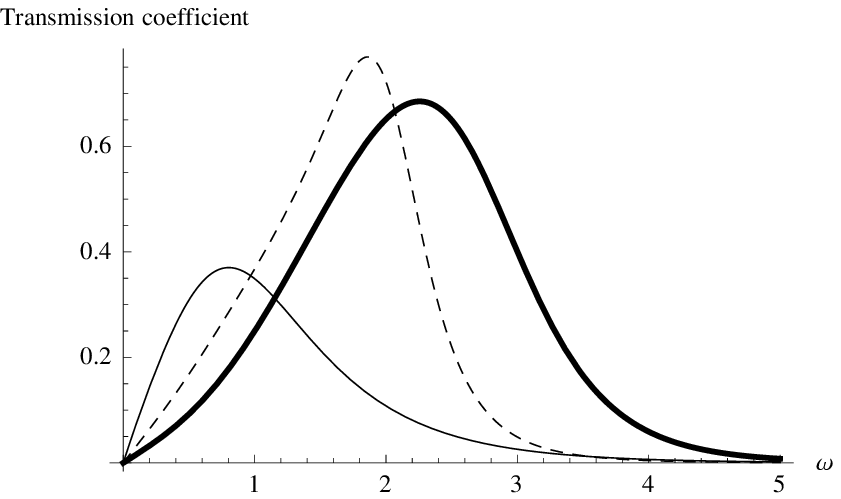}}\\
  \caption{\label{T} The behavior of the transmission coefficient with respect to $\omega$ for the following parameters $ M = 1 ; \; m^{2}l^{2}=-\frac{4}{5}; \; l=1; \; \kappa=1; h=-1$, for three different values $\xi=0$ (fine solid line); $\xi=0.1$ (dashed line) and $\xi=0.2$ (thick solid line)}
\end{figure}

\begin{figure}[h]
\centering
  \framebox{\includegraphics[width=4in,height=3in]{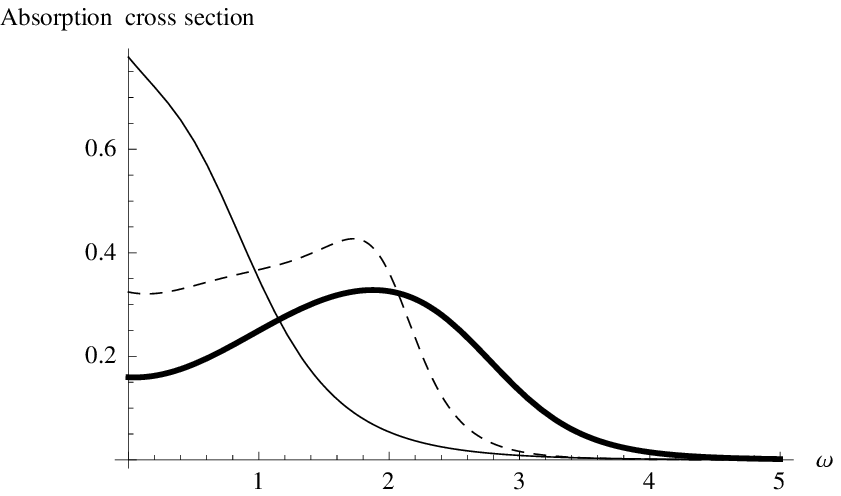}}\\
  \caption{\label{S} The behavior of the absorption cross section with respect to $\omega$ for the following parameters $M = 1 ; m^{2}l^{2}=-\frac{4}{5}; \; l=1; \; \kappa=1; h=-1$, for three different values $\xi=0$ (fine solid line); $\xi=0.1$ (dashed line) and $\xi=0.2$ (thick solid line)}
\end{figure}

The behavior of the reflection and absorption coefficients as well as the cross section are shown in Figures \ref{R}, \ref{T} and \ref{S} respectively.

It can be proven that the addition of the reflection and absorption coefficient adds up to $1$, which is consistent with the election of negative values for $h$.

\section{Lifshitz black hole quasinormal modes and their stability}

By assuming a null scalar field as boundary condition on the infinity, and without loss of generality choosing $\alpha = \alpha_- = - \frac{i}{2 M ^{3/2}} \, l \omega$, from equation (\ref{boundaryinfinity}) it can be seen that the field is regular in $r \rightarrow \infty$ for $\beta_- < 0$ if $ 1 + c + k + n = 0 $, for $\beta_+ > 2$ if $ 1 - c + s + n = 0 $, regarded $ \xi > - \frac{l^2 m^2}{ 26 }$. Using these conditions the quasinormal modes are obtained after lengthly algebra, however, as it was shown above, this problem is equivalent to the homogeneous one, the quasinormal modes are easily obtained by using the transformations (\ref{transformationfrombertha}), which yields
\begin{eqnarray}
\nonumber \omega_1 = & & \frac{2\,iM^{3/2}}{l} \{1 + 2n + \sqrt{ 4 + l^2 m^2 + 26 \xi } - [\frac{\kappa^2}{2M} + 7 + \frac{3 l^2 m^2}{2} + 6 n \left( n  + 1 \right) \\
\label{omega1a} &+& 35 \xi + 3 \left(1 + 2 n\right) \sqrt{ 4 + l^2 m^2 + 26 \xi}]^{1/2}\} \, ,
\end{eqnarray}
and
\begin{eqnarray}
\nonumber \omega_2 = & & \frac{2\,iM^{3/2}}{l} \{1 + 2n + \sqrt{ 4 + l^2 m^2 + 26 \xi } + [\frac{\kappa^2}{2M} + 7 + \frac{3 l^2 m^2}{2} + 6 n \left( n + 1 \right)\\
\label{omega1b} &+& 35 \xi + 3 \left(1 + 2 n\right) \sqrt{ 4 + l^2 m^2 + 26 \xi}]^{1/2}\} \, ,
\end{eqnarray}
where $n$ is a zero or a positive integer.

\begin{figure}[h]
\centering
  \framebox{\includegraphics[width=4in,height=3in]{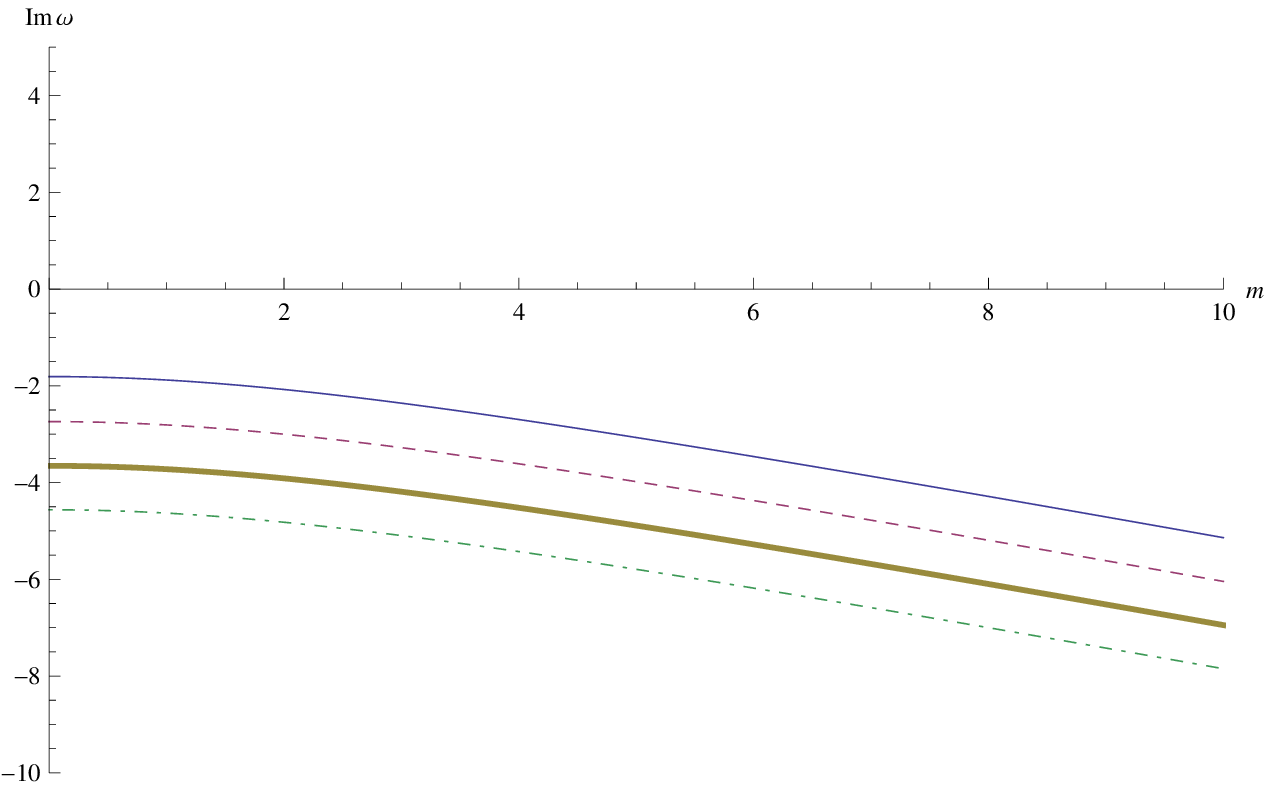}}\\
  \caption{\label{modos1} The behavior of the black hole's quasinormal modes (\ref{omega1a}) for the following parameters $\xi=\frac{1}{16}; \; m=1; \; l=1; \; M=1; \; \kappa=1$, where the thin solid line represents the mode $n=0$, the dashed line the mode $n=1$, the gross solid line the mode $n=2$ and the dashed/dotted line the mode $n=3$}
\end{figure}

\begin{figure}[h]
\centering
  \framebox{\includegraphics[width=4in,height=3in]{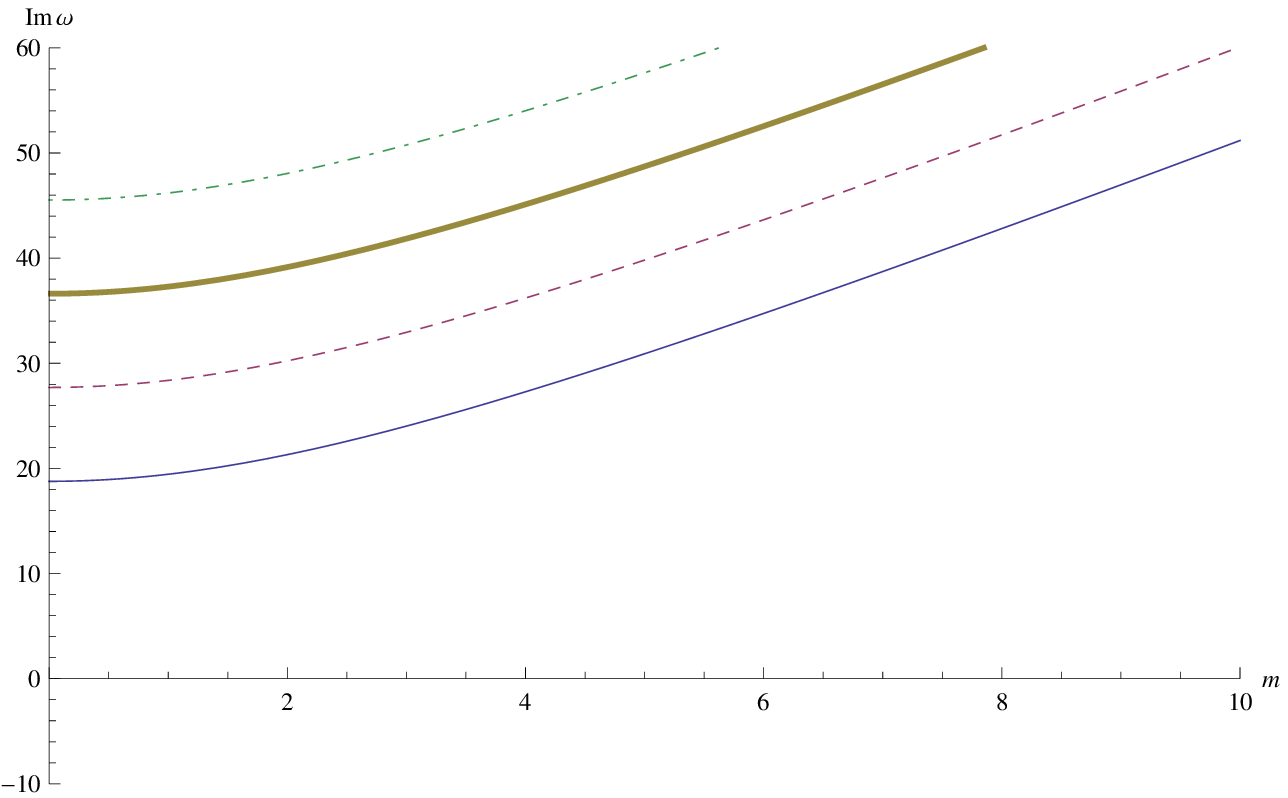}}\\
  \caption{\label{modos2}: Shows the behavior of the black hole quasi normal modes (\ref{omega1b}) for the following parameters $\xi=\frac{1}{16}; \; m=1; \; l=1; \; M=1; \; \kappa=1$, where the thin solid line represents the mode $n=0$, the dashed line the mode $n=1$, the gross solid line the mode $n=2$ and the dashed/dotted line the mode $n=3$}
\end{figure}

Stability of these solutions are restricted specifically by the term $\sqrt{ 4 + l^2 m^2 + 26 \xi}$, regarded $\mathfrak{Im}\{\omega\} \leq 0$. It is worth noting that this condition is completely equivalent to impose a restriction on the term of squared mass of the scalar field that we have used. To be more precise, if we define an effective mass on the following form
\begin{eqnarray*}
m^2_{effective} = m^2 + \frac{26 \, \xi}{l^2} \, ,
\end{eqnarray*}
this redefined effective mass must agreed with the analogue of the Breitenlohner-Freedman condition in order to have a stable propagation, which was established in \cite{Kachru}. Figure \ref{modos1}) shows that the first set of quasinormal modes are essentially stable, because of having a negative imaginary part. In fact, equation (\ref{omega1a}) is completely imaginary, therefore these modes have over damping state of oscillation. Figure \ref{modos2}) shows that the second set of quasinormal modes are instable because of having a positive imaginary part. In the same way than before equation (\ref{omega1b}) turns out to be completely imaginary, and so these modes are non convergent.

By using equation (\ref{solutionradialdiffx}) with $C_{2}=0$, due to the horizon boundary conditions, and using equation (\ref{F}), we are able to evaluate the flux at infinity in accord to the following identity regarding the derivative of the confluent Heun's functions \cite{Fiziev}
\begin{equation}\label{Heunderivative}
\frac{d}{dx}Heun_{C}(0,b,c,d,e;x)_{x=0}=\frac{1}{2}\left( \frac{b+c+bc+2e}{b+1}\right) \, ,
\end{equation}
which yields
\begin{equation}\label{asymptoticflux3}
F \left( x \rightarrow 1 \right) = \frac{2\rho _{+}^{4}}{l^{4}}\left\vert C_{1}\right\vert ^{2} \mathfrak{Im} \left( 2\left( \beta -1\right) B_{1}^{\ast }B_{2}+\left( \alpha - v_{1}\right) \left\vert B_{1}\right\vert ^{2}\left( 1-x\right) ^{2\beta - 1}+\left( \alpha -v_{2}\right) \left\vert B_{2}\right\vert ^{2}\left(1 - x\right) ^{3-2\beta }\right) \, ,
\end{equation}
where
\begin{eqnarray}
\label{v1} v_{1}=\frac{1}{2}\left( \frac{c+b+bc+2e+2d}{1+c}\right) \, , \\
\label{v2} v_{2}=\frac{1}{2}\left( \frac{-c+b-bc+2e+2d}{1-c}\right) \, ,
\end{eqnarray}
and $B_1$ and $B_2$ were defined on equations (\ref{b1} - \ref{b2}), where the flux was evaluated directly from the asymptotic radial equation. Now, imposing the boundary condition of vanishing flux at infinity, we are able to obtain two sets of quasinormal modes given by the conditions, $1-c+k+n=0 $ or $1+c+s+n=0$. One set is similar to the ones obtained from the Dirichlet boundary conditions of vanishing scalar field, however we find another set of quasi normal modes, that are stable for a range of imaginary scalar mass as it can be seen below. These modes are given by
\begin{eqnarray}
\nonumber \omega _{3} = & & \frac{ 2iM^{ 3/2 }}{ l } \{ \left( 1 + 2n - \sqrt{ 4 + l^{2} m^{2} + 26 \xi }\right) - [\frac{ \kappa^{2} }{2M} + 7 + \frac{ 3l^{2} m^{2}}{2} + 6n \left( n+1\right) \\
\label{omega2a} &+& 35\xi -3\left( 1+2n\right) \sqrt{4+l^{2}m^{2}+26\xi }]^{1/2}\} \, ,
\end{eqnarray}
and
\begin{eqnarray}
\nonumber \omega _{4} = & & \frac{ 2iM^{ 3/2 }}{l} \{ \left( 1 + 2n - \sqrt{ 4 + l^{2} m^{2} + 26 \xi }\right) + [ \frac{\kappa^{2} }{2M} + 7 + \frac{3l^{2}m^{2}}{2} + 6 n \left( n+1\right) \\
\label{omega2b} &+& 35\xi -3\left( 1+2n\right) \sqrt{4+l^{2}m^{2}+26\xi }]^{1/2}\}.
\end{eqnarray}

\begin{figure}[h]
\centering
  \framebox{\includegraphics[width=4in,height=3in]{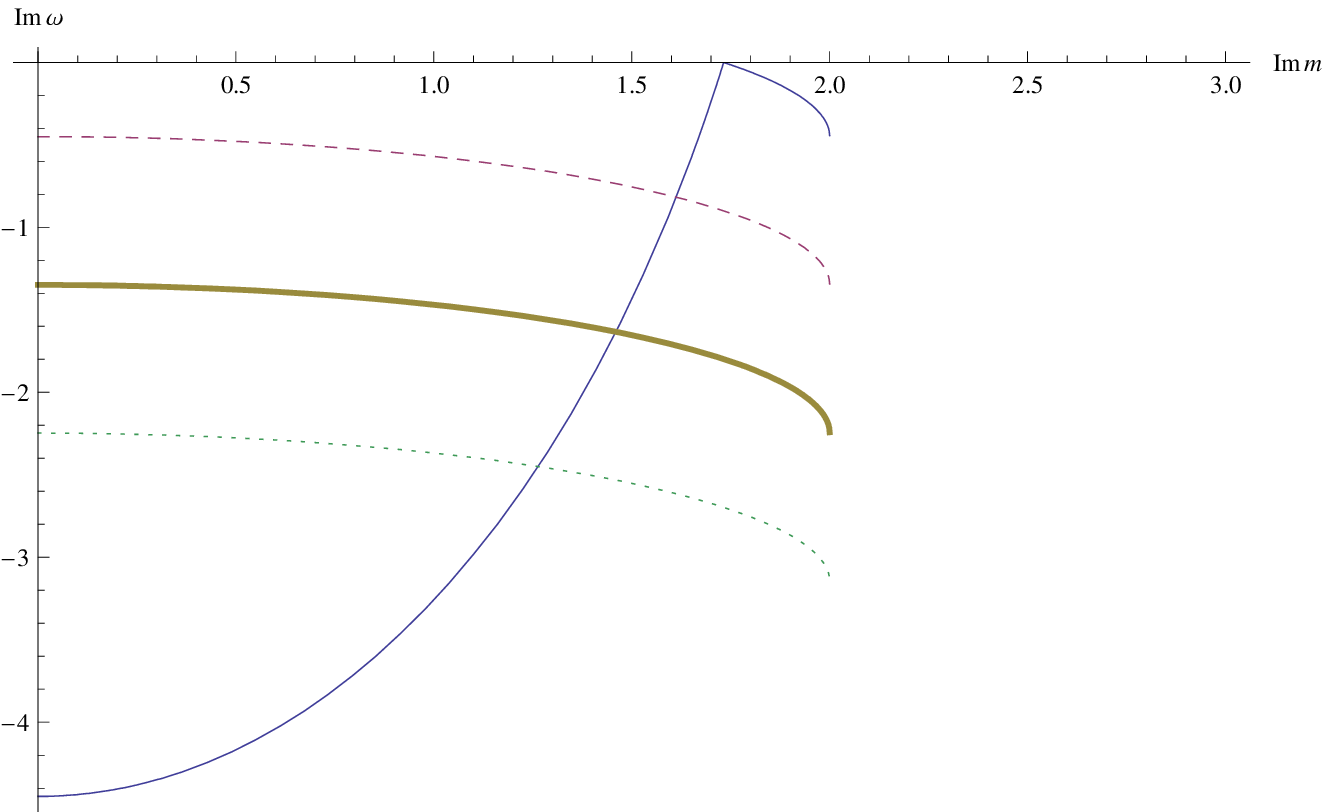}}\\
  \caption{\label{modos3} The behavior of the quasinormal modes (\ref{omega2a}) for the following parameters $\xi=0; \; m=1; \; l=1; \; \kappa=1$, where the thin solid line represents the mode $n=0$, the dashed line the mode $n=1$, the gross solid line the mode $n=2$ and the dashed/dotted line the mode $n=3$}
\end{figure}

From equation (\ref{asymptoticflux3}) and considering $\beta _{\pm }=1\pm \sqrt{1+\frac{m_{eff}^{2}l^{2}}{4}}$, where $\frac{m_{eff}^{2}l^{2}}{4}=\frac{m^{2}l^{2}}{4}+13\frac{\xi}{2}$ the allowable range for the parameters are
\begin{equation}\label{imaginarymasscase1}
1<\beta _{+}<2 \;\;\; \textit{and} \;\;\; 0<\beta _{-}<1 \, ,
\end{equation}
this sets the exponent of the first term on equation (\ref{asymptoticflux3}) yielding $2\beta -1 > 0 \, \, \rightarrow \,\, \beta > 1/2$ and $3-2 \beta > 0$ which corresponds to $\beta = \beta _{-}$, on the second term we have $\beta < 3/2 \, \rightarrow \, \beta =\beta _{+}$, from here we have
\begin{equation}\label{imaginarymasscase3}
\sqrt{3 + 26 \xi} < \mathfrak{Im} \{ m \, l \} < \sqrt{4 + 26 \xi} \, ,
\end{equation}
Figure \ref{modos3}) illustrates the fact that this last set of modes with imaginary mass are essentially stable.

The previous calculation is ensured by the following, consider the Klein-Gordon equation, which is essentially equation (\ref{radialdiffrho}) due to the fact that non-minimal coupling only redefines the constants $\kappa$ and $m$, this equation can be transformed into a Schr\"{o}dinger like equation by making use of the tortoise coordinate transformation, defined by
\begin{equation}\label{tortoise}
dx=\frac{1}{\left( \frac{\rho }{l}\right) ^{2}\left( \left( \frac{\rho }{l} \right) ^{2}-M\right) }d\rho \, ,
\end{equation}
then Klein-Gordon equation adopts the form
\begin{equation}\label{Schrodingerlike}
\left( \frac{d^{2}}{dx^{2}}+\omega ^{2}-V\left( \rho \right) \right) \left(\rho ^{1/2}R\left( x\right) \right) =0 \, ,
\end{equation}
where the effective potential can be identified with the following expression
\begin{equation}\label{effectivepotential}
V\left( \rho \right) =\frac{1}{l^{2}}\left( \frac{\rho }{l}\right)
^{2}\left( \left( \frac{\rho }{l}\right) ^{2}-M\right) \left( \kappa^{2}-8M\xi -\frac{3}{4}M-\left( m-\frac{26}{l^{2}} \xi -\frac{7}{4l^{2}}\right) \rho ^{2}\right) \, .
\end{equation}
Note that this potential is divergent on $\rho \rightarrow \infty$, therefore the condition of null flux at this boundary is justified.

\section{Soliton solution, quasinormal modes and its stability}

\begin{figure}[h]
\centering
  \framebox{\includegraphics[width=4in,height=3in]{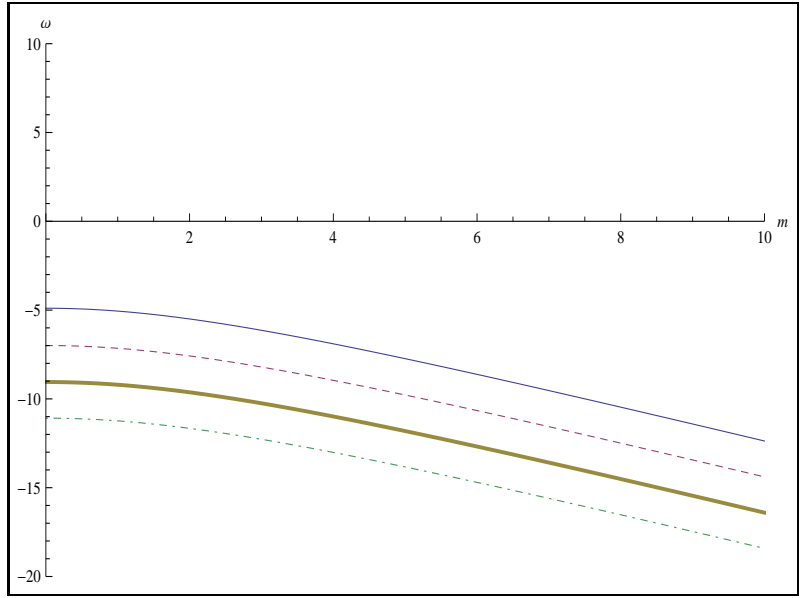}}\\
  \caption{\label{modos4}:Shows the behavior of the quasi normal soliton modes (\ref{omegasoliton1}) for the following parameters $\xi=\frac{1}{16}; \; l=1; \; \kappa_{soliton}=1$, where the thin solid line represents the mode $n=0$, the dashed line the mode $n=1$, the gross solid line the mode $n=2$ and the dashed/dotted line the mode $n=3$}
\end{figure}

A remarkable feature of the Lifshitz black hole is that by performing the double Wick rotation on the time and space coordinates
\begin{equation}\label{doubleWick}
\nonumber t \rightarrow - \frac{i \, l^4}{{\rho_+}^3} \bar \varphi; \;\;\;\;\; \varphi \rightarrow \frac{i \, l}{\rho_+} \bar t,
\end{equation}
and performing the change of variable
\begin{eqnarray} \label{changeofvariablesoliton}
\rho = \rho_+ \, \cosh \left( r \right),
\end{eqnarray}
the metric (\ref{lipshitzmetric}) is transformed into the following one:
\begin{equation} \label{solitonmetric}
ds^2 = -l^2 \, \cosh^2 \left( r \right) d\bar t^2 + l^2 dr^2 + l^2 \, \cosh^4 \left( r \right) \, \sinh^2 \left( r \right) d \bar \varphi^2,
\end{equation}
which can be recognized as a soliton metric \cite{soliton}.
Formally, the quasinormal modes of this metric must be obtained in the same way as was shown on the previous section. However, due to the symmetry of the transformations involved, the quasinormal modes for the soliton can be easily obtained by performing the following substitution
\begin{equation}\label{solitonsubstitution}
\nonumber \kappa \rightarrow - \frac{i \, \rho_+}{l} \omega_{soliton}; \;\;\;\;\; \omega \rightarrow \frac{i \, {\rho_+}^3}{l^4} \kappa_{soliton}.
\end{equation}
Inserting (\ref{solitonsubstitution}) in equations (\ref{omega1a}) and (\ref{omega1b}) yields
\begin{eqnarray}
\nonumber \omega_{soliton} = &\pm& \{ 4 + 2 \kappa_{soliton} - \frac{1}{2} \kappa^2_{soliton} +l^2 m^2 +4 n + 4 \kappa_{soliton} n \\
 \label{omegasoliton1} &+& 4 n^2 + 44 \xi + 2 \left( 1 + \kappa_{soliton} + 2n \right) \sqrt{4 + l^2 m^2 + 26 \xi} \}^{\frac{1}{2}}
\end{eqnarray}
Note that the stability of these solutions only depends on the term $\sqrt{4 + l^2 m^2 + 26 \xi}$ which it has been already restricted according to the to Breitenlohner-Freedman condition, the positivity or negativity of rest of the term in equation (\ref{omegasoliton1}) is not transcendent due to the fact that the 'minus' or 'plus' sign can be choose to ensure stability. The behavior of these modes can be seen on Figure \ref{modos4}) that shows explicitly the stability of the solutions presented.

In the same way, we find a second set of quasinormal modes for the soliton, by using the expressions found in equations (\ref{omega2a}) and (\ref{omega2b}), we have
\newpage
\begin{eqnarray}
\nonumber\omega _{soliton} = &\pm& \{ 4 + 2\kappa _{soliton}-\frac{1}{2}\kappa
_{soliton}^{2}+l^{2}m^{2}+4n+4\kappa _{soliton}n\\
\label{omegasoliton2} &+& 4 n^{2} + 44 \xi - 2 \left(1+\kappa _{soliton}+2n\right) \sqrt{4+l^{2}m^{2} + 26\xi }\}^{\frac{1}{2}}
\end{eqnarray}

\begin{figure}[h]
\centering
  \framebox{\includegraphics[width=4in,height=3in]{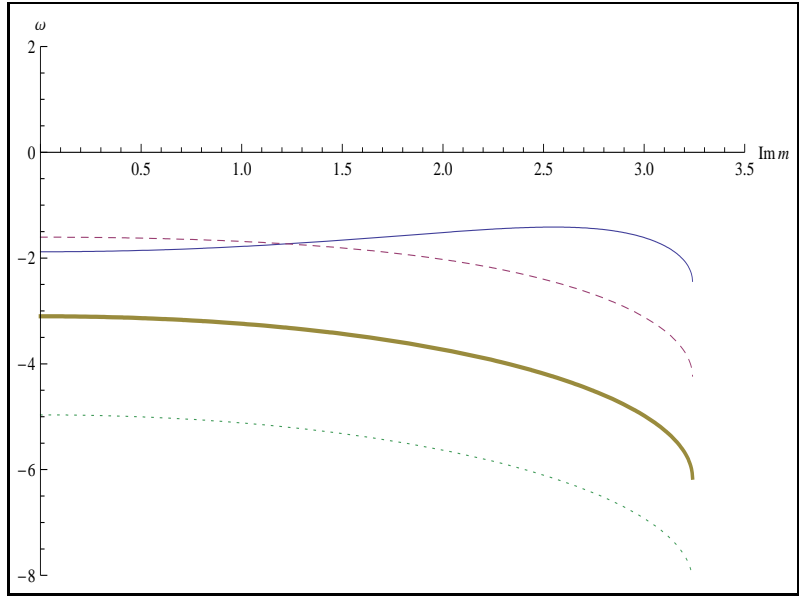}}\\
  \caption{\label{modos5} The behavior of the quasi normal soliton modes (\ref{omegasoliton2}) for the following parameters $\xi=\frac{1}{16}; \; l=1; \; \kappa_{soliton}=1$, where the thin solid line represents the mode $n=0$, the dashed line the mode $n=1$, the gross solid line the mode $n=2$ and the dashed/dotted line the mode $n=3$}
\end{figure}

The stability of this case depends on the condition $\frac{4 + 26 \xi}{l^2} > \| m \|^2$ which is valid for imaginary masses, recall that this is in accordance with the Breitenlohner-Freedman condition. As in the previous case, the stability is ensured by the suitable sign election of the term in the squared root. (\ref{omegasoliton2}). The behavior of these modes can be seen on Figure \ref{modos5}).

\section{Final Remarks}

We can summarize some important features found in this paper:

\begin{itemize}
\item The appearance of the coupling parameter explicitly in the absorption cross section (equation (\ref{crosssection}), imply that it is not possible to obtain the geometric area of the black hole as a result in the limit of low energies and for a s-wave, unless this parameter became null.

\item As can be seen from this study, non-minimal coupling doesn't affect the overall form of the quasinormal modes, moreover the non-minimal quasinormal modes can be easily obtained by redefining the parameters $\kappa$ and $m$. Although this can be considered a natural extension from the minimal problem, this is only possible by the simple form of the Ricci scalar in a Lifshitz background.

\item There is a complete new set of stables quasinormal modes that has been not considered before, these modes are associated to the possibility to have an imaginary mass for the scalar field. These modes are found by imposing the flux to be zero at infinity and requiring the expression to be valid according to the Breitenlohner-Freedman condition for stability .

\item By performing a double Wick rotation it is possible to change the metric in a way which is isomorphic to a soliton type of metric, and from here is easily seen that the quasinormal modes for this metric can be obtained from the black hole quasinormal modes by a simple transformation. These modes turn out to be
    stables.
\end{itemize}

\section{Acknowledgements}

This research was supported by VRIEA-DI-037.419/2012, Pontificia Universidad Cat\'olica de Valpara\'iso (SL), Fondecyt 1110076 (SL), DI12-0006 of Direcci\'on de Investigaci\'on y Desarrollo, Universidad de La Frontera (FP) and DI11-0071 Direcci\'on de Investigaci\'on y Desarrollo, Universidad de La Frontera (YV).

\section*{References}
\bibliographystyle{elsarticle-num}
\biboptions{comma,square}
\bibliography{./biblio}
\end{document}